# Epitaxial CeO$_2$ Films as a Host for Quantum Applications

Pralay Paul[1,2,*], Kusal M. Abeywickrama[1,2], Nisha Geng[3], Mritunjaya Parashar[4], Levi Brown[5], Mohin Sharma[4], Darshpreet Kaur Saini[4], Melissa Ayala Artola[1,2], Todd A. Byers[4], Bibhudutta Rout[4], Yiwei Ju[5], Xiaoqing Pan[6], Sumit Goswami[7], Sreehari Puthan Purayil[1,2], Casey Kerr[1,2], Dhiman Biswas[1,2], Ben Summers[1,2], Bin Wang[8], Horst Hahn[9], Alisa Javadi[1,2,10,*] and T. Venkatesan[1,2,10,*]

[1]Center for Quantum Research and Technology, University of Oklahoma, OK 73019

[2]Homer L Dodge Department of Physics and Astronomy, University of Oklahoma, OK 73019

[3]Sustainable Chemical, biological and materials engineering, University of Oklahoma, OK 73019

[4]Ion Beam Laboratory, Department of Physics, University of North Texas, Denton TX 76203

[5]Department of Physics and Astronomy, University of California Irvine, Irvine CA 92697

[6]Materials Science and Engineering Department, University of California Irvine, Irvine CA 92697

[7]MREC, University of Oklahoma OK 73019

[8]Department of Chemical and Biological Engineering, Tufts University, Boston MA

[9]Department of Materials Science, University of Arizona, Tucson AZ 85721

[10]School of Electrical and Computer Engineering, University of Oklahoma, OK 73019

## Abstract

In highly purified host, the coherence of quantum emitters is ultimately limited by hyperfine interactions between the emitter and lattice nuclei possessing non-zero nuclear magnetic moments. This limitation can only be mitigated through isotopic purification. In this work, we investigate CeO$_2$ as a host composed entirely of nuclei with zero nuclear moment. High-quality CeO$_2$ thin films were grown by PLD and doped with Tm and Er ions. Structural characterization using X-ray diffraction, atomic force microscopy, and ion channeling confirms single-crystalline, atomically smooth films with dopants substitutionally incorporated at Ce lattice sites. Photoluminescence lifetime measurements show significantly longer lifetimes for Er-doped CeO$_2$ (2.9 - 5.3 ms) compared with Tm-doped films (14 - 68 μs). Moreover, the Er-doped PLD films exhibit longer lifetimes at ~1% dopant concentration than previously reported for MBE-grown films. Density functional theory calculations reveal a substantial overlap between unoccupied O 2p and Tm 4f states near the valence band maximum, whereas Er 4f states remain well isolated. This electronic interaction likely introduces non-radiative recombination pathways in Tm-doped CeO$_2$, explaining the reduced lifetimes. These findings highlight the importance of selecting appropriate dopant-host combinations and optimized growth conditions to minimize non-radiative channels for quantum applications.

## I. Introduction:

Currently, many quantum emitters have been identified in a diverse set of host matrices, including nitrogen-vacancy [1, 2] and silicon-vacancy (SiV) [3-5] centers in diamond, divacancies and silicon vacancies in silicon carbide (SiC) [6, 7], or color centers in two-dimensional materials [8]. The presence of isotopes with non-zero nuclear spin, such as $^{13}$C (1.1%) in diamond or $^{29}$Si (4.7%) in SiC, introduces spin bath decoherence via hyperfine interaction [9]. While impressive spin coherence times, longer than one millisecond, have been observed in these materials, such demonstrations typically require isotopically purified hosts to remove the decoherence stemming from nuclei with non-zero magnetic moment.

Recently, wide bandgap oxides have emerged as promising hosts for quantum emitters [10-13]. These materials allow for cost-effective single-crystal growth and offer a broad choice of lattice-matched substrates. Among various strategies, doping with rare-earth (RE) ions such as $Er^{3+}$, $Eu^{3+}$, and $Tm^{3+}$ has gained attention due to the characteristic 4f-4f optical transitions, which are well shielded by the outer $5s^2$ and $5p^6$ shells [14-17]. This shielding renders transitions relatively immune to perturbations from the host lattice, enabling long radiative lifetimes and high optical coherence. For example, $Er^{3+}$ ions in $Y_2SiO_5$ exhibit radiative lifetimes on the order of ~ 11 ms [18], while $Eu^{3+}$ in $Y_2O_3$ demonstrates lifetimes around 0.8 ms [19]. Other notable systems include $Pr^{3+}$ in $YAlO_3$ [20] and $Nd^{3+}$ in $LaF_3$ [21], both of which exhibit narrow linewidths and long coherence under cryogenic conditions. Today, various deposition processes like molecular beam epitaxy, pulsed-laser deposition (PLD), and sputtering have demonstrated the capability to grow such oxides with atomic-level control.

Oxygen is nearly void of magnetic nuclei, which puts oxides at an advantage as a host with enhanced spin coherence. In particular, Cerium oxide ($CeO_2$) offers a unique advantage as all of cerium's stable isotopes have zero nuclear spin, and oxygen's only spin-active isotope ($^{17}$O) has a natural abundance of just 0.04%, making $CeO_2$ effectively a magnetically purified host [22-24]. This feature significantly suppresses spin bath-induced decoherence, preserving the intrinsic coherence of rare-earth dopants and potentially color centers. Indeed, $Er^{3+}$ doped in $CeO_2$ has demonstrated optical lifetimes as high as 1.5 ms at 4 K [25, 26], double that of $Eu^{3+}$ in $Y_2O_3$ [19], highlighting the potential of $CeO_2$ for rare earth-based quantum emitters. Furthermore, large-area single-crystalline $CeO_2$ can be grown on a range of substrates, including silicon.

Among the rare-earth dopants, thulium ($Tm^{3+}$) is a compelling candidate. Its strong $^3H_4 \rightarrow {^3H_6}$ emission at ~ 794 nm [27] lies within the first biological transparency window and the near-infrared (NIR) telecommunication band. In addition, $Tm^{3+}$ ions have an ionic radius (88 pm) closely matching that of $Ce^{4+}$ (87 pm), facilitating efficient substitution at the Ce Wyckoff sites in the fluorite $CeO_2$ lattice.

In this work, we present a comprehensive structural and optical investigation of epitaxial $Tm^{3+}$ and $Er^{3+}$ doped $CeO_2$ thin films grown by PLD on Si (001) and YSZ (001) substrates. High-resolution x-ray structural analysis confirms epitaxial growth of $CeO_2$ with fluorite symmetry, and MeV helium ion Rutherford backscattering and channeling measurements indicate near-perfect incorporation of Tm ions in $CeO_2$ lattice. Photoluminescence spectroscopy reveals strong NIR emission centered around 793 nm and 808 nm, with comparable intensities at room temperature and 5 K, indicating thermal stability of the emission process. Notably, we observe up-conversion of NIR excitation (1150-1207 nm) in resonance with the transitions of thulium ions (794 and 808nm). We have also studied the dynamics of the photoluminescence and have compared the lifetimes observed in Tm with that of Er in films prepared by PLD. To further elucidate the lifetimes observed in Tm with respect to Er we have performed DFT calculations to explore oxygen hybridization with the rare earth to understand non-radiative pathways.

## II. Experimental Details

### A. Growth of epitaxial films of $RE^{3+}$: $CeO_2$

High-density ceramic targets of $CeO_2$, $Ce_{0.99}Er_{0.01}O_3$, and $Ce_{0.9}Tm_{0.1}O_3$ were synthesized from high-purity precursors of $CeO_2$ (99.999%), $Er_2O_3$, and $Tm_2O_3$ (99.999%), procured from Sigma-Aldrich. The powders were homogeneously mixed, compacted using a cold isostatic press (CIP), and subsequently sintered at 1300° C to achieve dense, crack-free 1 inch diameter targets. Epitaxial thin films of $Tm^{3+}$-doped $CeO_2$, with Tm concentrations of 10, and 1 at.% were deposited on single-crystal YSZ (001) and Si (001) substrates using pulsed laser deposition (PLD, Neocera) in an ultra-high vacuum (UHV) chamber with in-situ reflection high-energy electron diffraction (RHEED) monitoring. Prior to deposition, YSZ (001) and Si (001) substrates were cleaned sequentially with deionized (DI) water, acetone, and isopropanol (IPA), followed by a final rinse with DI water to eliminate organic residues and particulates. The cleaned YSZ (001) substrates were then annealed in ambient air at 1000° C for 3 hours to enhance surface crystallinity and morphological uniformity. On the other hand, single-side polished Si (001) substrates with native $SiO_2$ surface layers were introduced into the PLD chamber and subjected to a tailored deposition protocol to accommodate the amorphous interfacial oxide. Once the base pressure in the chamber was reduced significantly below $1\times10^{-5}$ Torr (a reducing condition), the substrates were gradually heated to the final deposition temperature. An initial buffer layer of ~ 7 nm $CeO_2$ was deposited at a repetition rate of 2 Hz in vacuum, serving to facilitate epitaxial alignment along $CeO_2$ (111) by sequential removal of the native oxide layer. This was followed by the growth of 500 nm $CeO_2$ (or Tm-doped $CeO_2$ using a multi-target carrousel) under a controlled oxygen partial pressure ($1.1\times10^{-3}$ Torr). In contrast, depositions on single-crystal YSZ (001) substrates were performed directly under the desired oxygen ambience, without the buffer layer, owing to the favorable lattice match and absence of an interfacial oxide layer. This facilitates to grow $CeO_2$ (or Tm:$CeO_2$) film along (001) orientation. Epitaxial thin films of $Er^{3+}$-doped $CeO_2$ with Er concentration of 1 at.% were deposited on Si (001) and YSZ (001) substrates following the protocols discussed above. A KrF excimer laser (λ = 248 nm) was employed for ablation, with the laser fluence maintained at ~ 1.5 J/cm² and a target-to-substrate distance of ~ 5 cm. The depositions were performed at a substrate temperature of ~ 820° C under an oxygen partial pressure of $1.1 \times 10^{-3}$ Torr, followed by a 30-minute annealing at the same temperature with the oxygen pressure remaining unchanged during the annealing and cooling process.

### B. Structural, crystalline quality characterization

X-ray diffraction (XRD) and X-ray reflectivity (XRR) measurements were conducted using a Rigaku SmartLab diffractometer equipped with a Cu K$_α$ radiation source (λ = 1.5406 Å), operating at an accelerating voltage of 45 kV and a current of 200 mA. High-resolution XRD scans were performed in the 2θ-ω configuration to evaluate the crystallographic structure, phase purity, and out-of-plane orientation of the deposited thin films. XRR measurements were employed to determine film thickness, and interfacial roughness, utilizing a reflectivity mode with an angular resolution finer than 0.005°, ensuring precise quantification of the film structure and surface morphology.

Rutherford Backscattering Spectrometry (RBS) and ion channeling (RBS/C) measurements were conducted using a NEC 9SH 3MV Pelletron accelerator at the Ion Beam Laboratory (IBL), University of North Texas (UNT) [28]. The details of the measurement are in supplementary information (S.I.) section S1.

For scanning transmission electron microscopy (STEM) and energy-dispersive X-ray spectroscopy (EDS) analysis, cross-sectional lamellas of the thin film samples were prepared in the Tescan dual beam SEM-FIB via FIB lift-out method. To enable atomic-resolution analysis, the lamellas were further thinned to ~ 30 nm using a Fischione 1040 Nanomill. STEM imaging and EDS characterization were performed on a JEOL JEM-ARM300F TEM, operated at 300 kV with aberration correction. For this study the probe size was calibrated to 0.09 nm, with a convergence semi-angle of 24 mrad and 68 pA probe current. High-resolution STEM images were collected using a high-angle annular dark field (HAADF) detector (42 mrad -180 mrad). For atomic-resolution EDS mapping, a scan area of < 4 × 4 nm² with < 5 Å step size and 0.1 ms dwell time was used, with drift correction applied after each scan.

### C. Computational Methods

Geometry optimizations and the densities of states (DOS) were carried out using density functional theory (DFT) as implemented in the Vienna ab-initio Simulation Package (VASP) version 6.5 [29, 30], with the projector augmented wave (PAW) method [31]. The generalized gradient-corrected exchange and correlation functional of Perdew-Burke-Ernzerhof (PBE) [32] was employed, and the Ce $4f^1 5s^2 5p^6 5d^1 6s^2$, O $2s^2 2p^4$, Tm $4f^{13} 5s^2 5p^6 6s^2$, and Er $4f^{12} 5s^2 5p^6 6s^2$ electrons were treated explicitly in all of the calculations. A simple rotationally invariant approach [33] was considered with an effective Hubbard U parameter of U = 5.0 eV for Ce [34], and U = 7.3 eV [35-37] for Tm and Er to account for localized Ce and Tm 4f electrons, as recommended in the literature. The plane-wave basis set energy cutoffs were 700 eV in DFT+U calculations, and the energy converged within 1 meV/atom. All spin-polarized calculations were performed in a 2x2x2 supercell of $Fm\bar{3}m$ $CeO_2$ containing 96 atoms using a 3x3x3 Γ-centered Monkhorst-Pack k-point mesh [38]. Magnetic moments were obtained from spin-polarized calculations.

## III. Results and Discussion
### A. Structural and crystalline quality of the $RE^{3+}$: $CeO_2$ films

To achieve epitaxial growth and maintain stoichiometry in $Tm^{3+}$: $CeO_2$ thin films, key deposition parameters including substrate temperature, target to substrate distance, laser fluence, and partial oxygen pressure ($P_{O2}$) were systematically optimized. Figure 1a presents the high-resolution X-ray diffraction (HRXRD) 2θ-ω scan of a 10 at.% Tm-doped $CeO_2$ film grown on YSZ (001) substrate using these optimized conditions. Whereas HRXRD pattern of other films grown on both YSZ (001) and Si (001), following the same growth parameters, are shown in the S.I. (Figure S1-S4). The HRXRD pattern unequivocally confirms the formation of a single-phase fluorite-structured Tm and Er-doped $CeO_2$ as shown in Figure 1a and Figure S4. In Figure 1a, the diffraction peaks observed at ~ 33.06° and ~ 69.42° correspond to the (002) and (004) reflections of Tm-doped $CeO_2$, respectively. This yielded an average out-of-plane lattice constant for film of ~ 5.405 Å. Based on the cubic lattice parameters of YSZ (~ 5.14 Å) and the film (~ 5.405 Å), the in-plane lattice mismatch is estimated to be ~ 5.16%, indicating that the film experiences compressive strain when coherently grown on the YSZ substrate. No noticeable shift in the $CeO_2$ peak position is observed in any of the samples, as the ionic radii of Tm and Ce are close.

The inset of Figure 1a highlights a magnified view around the (002) reflection of the film, revealing well-defined Kiessig (thickness) fringes. The presence of these interference fringes is indicative of coherent layer-by-layer growth and underscores the high structural quality of the film. The film thickness estimated from the fringe periodicity is ~ 37 nm, in excellent agreement with values obtained from both XRR and RBS. To further evaluate crystalline quality, rocking curve (RC) measurements were performed around the (002) reflections of both the YSZ substrate and the Tm: $CeO_2$ film. The substrate exhibited a sharp RC with a full width at half maximum (FWHM) of ~ 0.01°, indicative of its excellent crystallinity. The RC of the film displayed a composite profile featuring a sharp component atop a broader base. This suggests a strained interface region within the initial few nanometers of the film, yielding a sharp RC peak (FWHM ~ 0.04°). Beyond a critical thickness, the film relaxes (due to large lattice mismatch), resulting in a broader RC component (FWHM ~ 0.6°), while maintaining its epitaxial orientation along the (001) direction. A similar observation was reported by Yezhou Shi et al. [39] in highly strained $CeO_2$ films grown on YSZ.

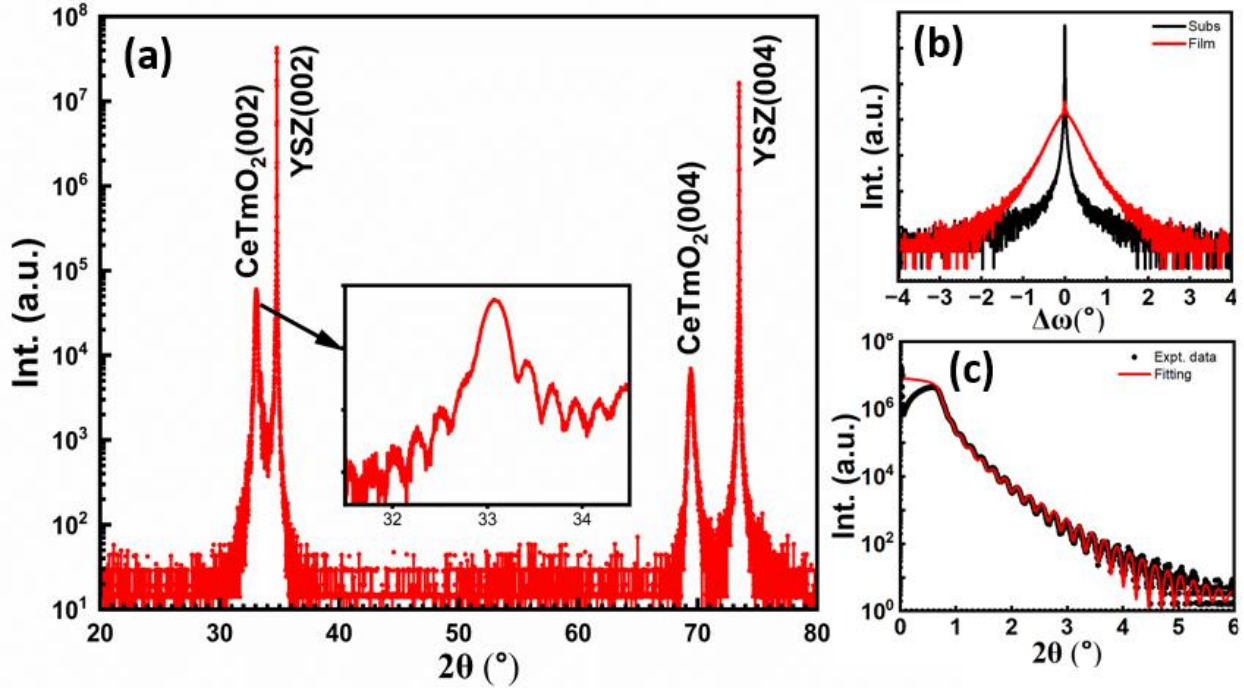

**Figure 1: (a)** 2θ-ω spectrum of 10% Tm-doped $CeO_2$ (001) grown epitaxially on YSZ (001). **(b)** shows the Rocking Curve measured around (002) peaks of substrate (black color) and film (red color), and **(c)** shows X-ray reflectivity data of the film (black color) for thickness optimization along with fitting (red color).

The XRR measurement and corresponding fitting results, shown in Figure 1c, further confirm a film thickness of ~ 38 nm, consistent with the value derived from HRXRD interference fringes. The XRR derived surface roughness was estimated to be ~ 0.6 nm, indicative of an atomically smooth film surface. This result was independently corroborated via atomic force microscopy (AFM), as shown in the S. I. (Figure S5), which yielded a comparable root mean square surface roughness of ~ 0.5 nm.

Figure S3 shows the 2θ - ω X-ray diffraction scan of the 10 at.% Tm-doped $CeO_2$ thin film grown on a Si (001) substrate. The film exhibits a strong preferential orientation along the (111) direction. Distinct diffraction peaks corresponding to the (111) and (222) planes are observed at ~ 28.8° and ~ 59.7°, respectively, yielding an average out-of-plane lattice parameter of ~ 5.35 Å for the Tm:$CeO_2$ film. The stabilization of the (111) orientation on Si (001) is likely governed by the lower surface energy of the $CeO_2$ (111) plane [40]. Considering the in-plane lattice constant of Si (~ 5.431 Å), the film exhibits an in-plane lattice mismatch of ~ −1.56%, indicating tensile strain in the film. Rocking curve measurements of the $CeO_2$ (111) reflection show a broader FWHM of ~ 4.2°, whereas the Si (004) substrate peak exhibits a significantly narrower FWHM of ~ 0.002°. Structural analysis indicates that films grown on YSZ (001) substrates possess superior crystalline quality compared to those grown on Si (001). The use of Si (001) substrates in this study was primarily intended to independently verify the photoluminescence emission from the Tm:$CeO_2$ film, since YSZ substrates contain rare-earth Y ions that may exhibit optical transitions similar to Tm, whereas Si does not contribute to any such rare-earth-related emission.

Figures 2a and 2b present the Rutherford backscattering (RBS) spectra and the corresponding elemental concentrations for a CeO$_2$ sample doped with 10 at.% Tm. The fitting of the RBS data confirms a Tm concentration of 10 at%, consistent with the intended Ce$_{0.9}$Tm$_{0.1}$O$_2$ stoichiometry based on the growth parameters. Figure 2c shows the random and channeled yield for the Ce$_{0.9}$Tm$_{0.1}$O$_2$/YSZ(001) sample, showing a significant decrease in the ion channeling yield with a $\chi_{min}$ of ~ 5.49% and ~ 3.09% for Tm and Ce atoms, suggesting excellent crystallinity of the sample. It is worth noting that the YSZ substrate exhibited a reduced $\chi_{min}$, although the value remained relatively high at approximately 26.99%. This increase in substrate $\chi_{min}$ is primarily attributed to beam dechanneling as it traverses the top Ce$_{0.9}$Tm$_{0.1}$O$_2$ layer. Such dechanneling is due to energy loss during interactions with near-surface atoms, and even minor disturbances in the beam's path can cause its profile to spread as it penetrates deeper into the target. Figure 2d presents the angular yield curves recorded at various $\theta_1$ angles, revealing that the profiles for YSZ, Ce, and Tm are nearly identical, with their respective $\chi_{min}$ values all occurring at $\theta_1 = 0°$. The $\Psi_{1/2}$ values were also closely matched and found to be 0.80° for Ce and 0.90° for Tm, indicating that Tm atoms are successfully occupying regular Ce lattice positions. While dechanneling caused the $\Psi_{1/2}$ for YSZ to differ slightly, around 0.56°, the centroid of the angle still aligns with those of Ce and Tm, implying good lattice matching between the YSZ substrate and the deposited Ce$_{0.9}$Tm$_{0.1}$O$_2$ film. Using the $\chi_{min}$ values derived from Figure 2d, the substitutional fraction, $f_s$, for Tm atoms was calculated to be ~ 98.22%, indicating that most Tm atoms are incorporated at the regular Ce lattice sites.

Further verification of Tm dopant incorporation into the CeO$_2$ thin film was obtained via high-resolution STEM analysis, wherein the atomic structure of the lattice is directly imaged. Figure 3a presents a low-magnification cross-sectional STEM-HAADF image of the 10 at. % Tm doped CeO$_2$ film along [100] zone axis. The dark region within the CeO$_2$ film arises from thickness variations introduced during FIB thinning, rather than from intrinsic defects, and does not affect the evident single-crystalline quality of the film. Figure 3b shows a high-magnification STEM-HAADF image of the area highlighted in red in Figure 3a. Corresponding EDS elemental maps of Ce, O, and Tm are displayed in Figures 3c-e, respectively. EDS maps of Ce and O show their respective atomic lattices, while the map of Tm confirms homogeneous substitutional doping throughout the CeO$_2$ lattice. The STEM results align well with RBS and XRR measurements, which estimate the film thickness at approximately 35 nm.

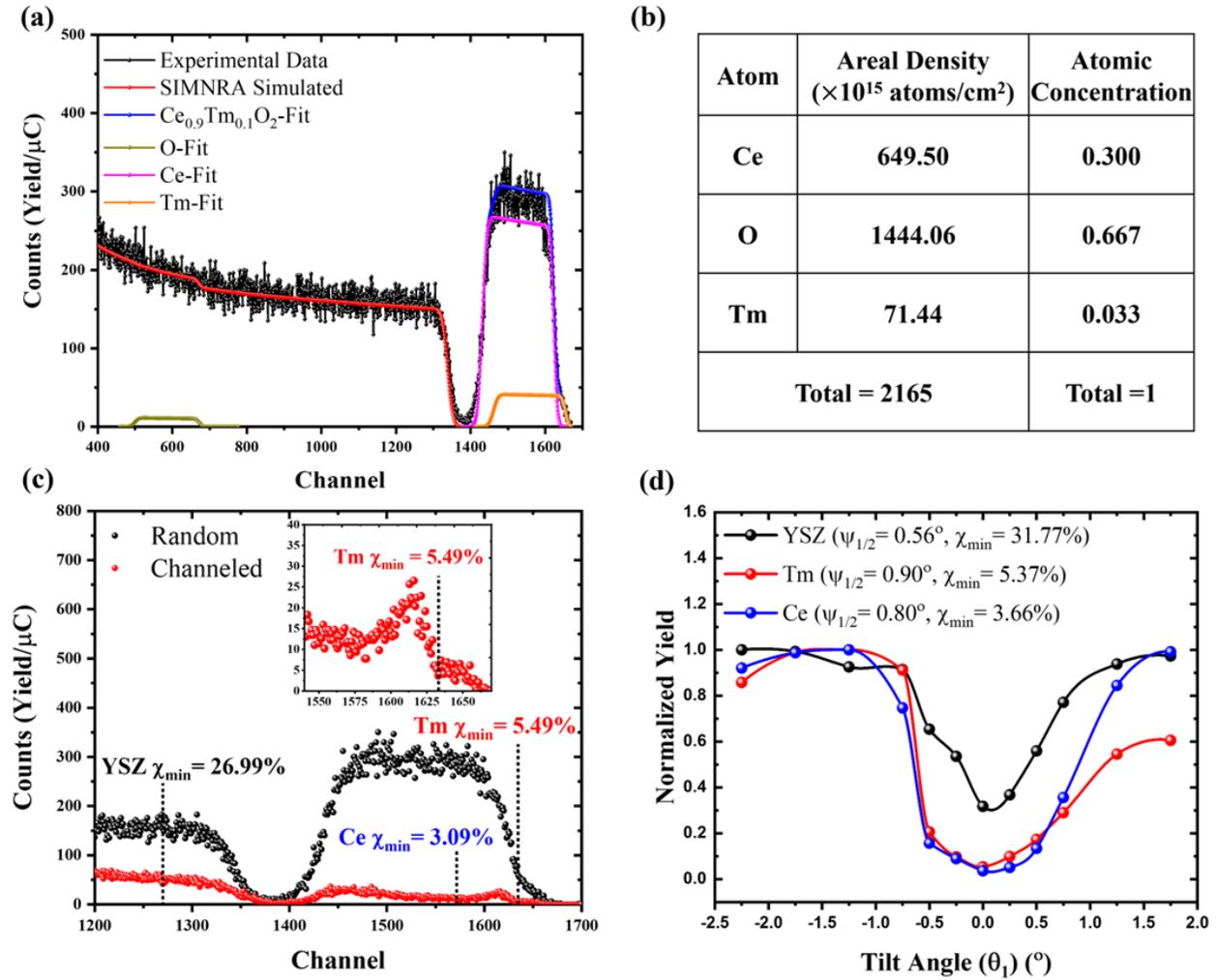

**Figure 2: (a)** Experimental and simulated RBS spectrum of $Ce_{0.9}Tm_{0.1}O_2$/YSZ sample showing individual contributions from the Ce, O, and Tm species. **(b)** The areal density and atomic concentration of the Ce, O, and Tm species derived from the RBS results. **(c)** RBS spectra in random and channeling orientation showing a comparable and significant drop in the Ce ($\chi_{min}$ = 5.49%) and Tm ($\chi_{min}$ = 3.09%) yield, indicating excellent crystallinity of the $Ce_{0.9}Tm_{0.1}O_2$ layer. (Inset highlights the Tm signal in the spectrum) **(d)** Angular yield curves for YSZ, Tm, and Ce obtained at different tilt angles ($\theta_1$) from the normal incidence. The angular curves show that the $\chi_{min}$ for both Ce and Tm occur at $\theta_1 = 0°$, with closely aligned profiles. The $\Psi_{1/2}$ values for Ce and Tm are 0.80° and 0.90°, respectively, and a calculated substitutional fraction ($f_s$) for Tm is 98.22%, supporting the conclusion that Tm atoms are effectively substituted into Ce lattice sites. (Note: Slight discrepancies in $\chi_{min}$ between Figures 2c and 2d may arise from factors such as damage accumulation from the $He^+$ beam or minor positioning errors from the goniometer.)

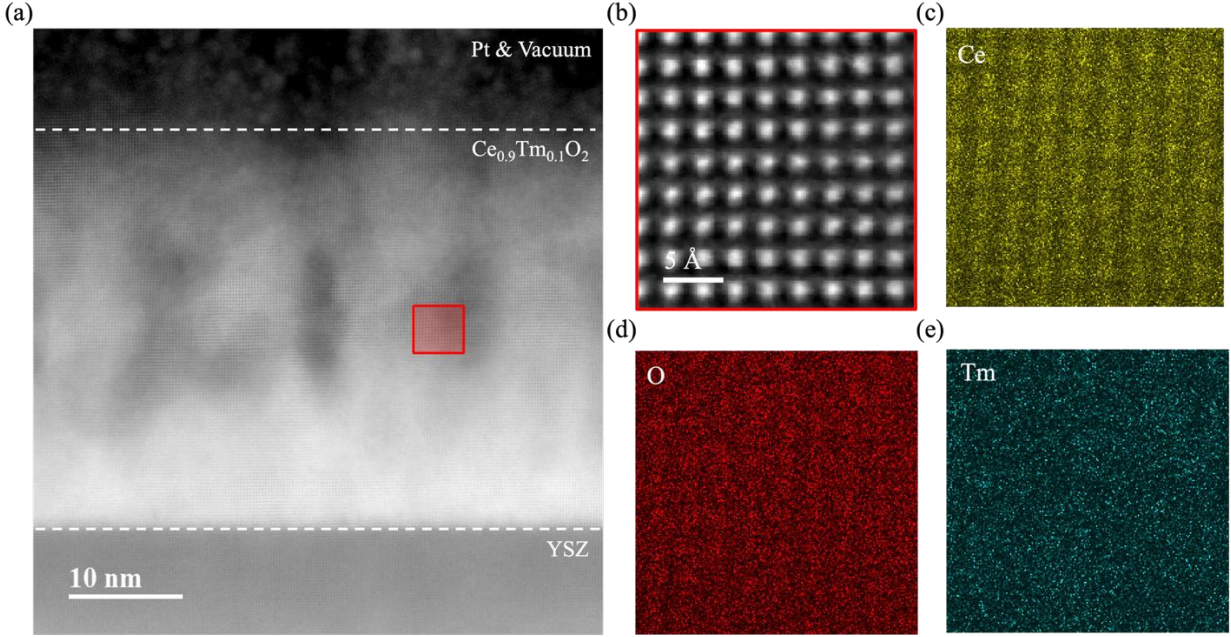

**Figure 3: STEM results of $Ce_{0.9}Tm_{0.1}O_2$/YSZ thin film**. (a) Low magnification STEM-HAADF image of cross-sectional $Ce_{0.9}Tm_{0.1}O_2$/YSZ thin film. (b) High-magnification STEM-HAADF image of the red-highlighted region in (a). (c) EDS mapping of Ce L edge signal. (d) EDS mapping of O K edge signal. (e) EDS mapping of Tm M edge signal.

## B. Optical Characterization

Figure 4a presents the NIR PL spectra obtained from a $CeO_2$(200nm)/$Ce_{0.99}Tm_{0.01}O_2$(150nm)/$CeO_2$(200nm) grown on both YSZ (001) and Si (001) substrates under optimal conditions. The spectra were recorded under an excitation wavelength of 355 nm. Cerium oxide ($CeO_2$) is a wide-bandgap semiconductor with an optical bandgap of approximately 3.2 eV [41], corresponding to an excitation wavelength near 380 nm and hence the excitation wavelength lies well above the bandgap. The NIR emission lines can be attributed to radiative transitions within the $Tm^{3+}$ ion. The prominent emission at ~793 nm arises from the $^3H_4 \rightarrow ^3H_6$ transition, which shows the highest intensity under 355 nm excitation. A comparably strong band near 808 nm is also observed; we attribute this feature to crystal-field induced Stark splitting within the $Tm^{3+}$ manifolds rather than assigning it to a distinct electronic transition. The excitation mechanism has been understood as follows: photoexcitation at 355 nm promotes electrons from the Oxygen $2p$ valence band to the Ce $4f$ conduction states, forming a charge transfer state. Energy is then non-radiatively transferred from this charge transfer state to the $^1G_4$ level of $Tm^{3+}$, followed by radiative relaxation through intra-$4f$ transitions [42]. To definitively attribute the observed emission to the doped films rather than substrate contributions, additional PL measurements were performed on substrates without films and pure $CeO_2$ films which showed no intrinsic emission in this spectral range shown as shown in the S.I. (Figure S6-S9).

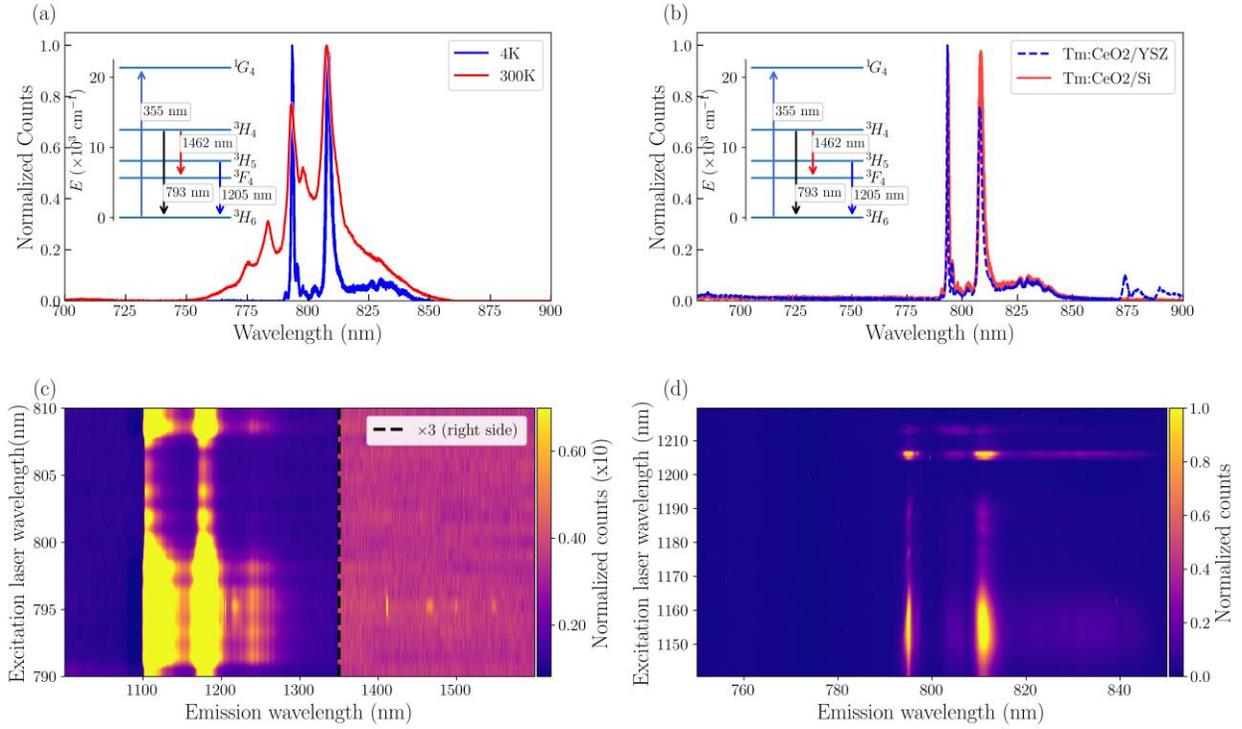

**Figure 4: (a)** Temperature-dependent PL spectra of 1 at.% $Tm^{3+}$:$CeO_2$/Si film excited at 355 nm, showing the thermal quenching behavior of the $^3H_4 \rightarrow {}^3H_6$ transition near 800 nm. **(b)** Photoluminescence (PL) spectra of 1 at.% $Tm^{3+}$:$CeO_2$ films grown on YSZ(001) (blue) and Si(001) (red) substrates with a 355 nm laser. The inset (identical in both PL panels) illustrates the $Tm^{3+}$ energy levels involved in the excitation and emission processes. **(c)** Photoluminescence excitation (PLE) map of 1 at.% $Tm^{3+}$:$CeO_2$/Si film obtained by scanning the excitation laser across the $^3H_4 \rightarrow {}^3H_6$ transition region while monitoring longer-wavelength emissions. The colormap is intentionally saturated to highlight $Tm^{3+}$ emission features. To enhance visibility of weaker spectral features at longer emission wavelengths, the signal on the right side of the map ($\lambda > 1350$ nm) is multiplied by a factor of 3. A vertical dashed line indicates the boundary where this scaling is applied. **(d)** Up-conversion emission from $Tm^{3+}$ ions in $CeO_2$ under near-infrared excitation: a pulsed laser is tuned from 1170 nm to 1250 nm, and visible/near-infrared emission is recorded. A strong up-conversion response is observed when the excitation is resonant with the transition near 1210 nm.

A comparison between the room-temperature and low-temperature PL spectra (Figure 4b) reveals that cooling significantly sharpens the emission lines and enhances their relative intensity contrast. At 4 K, the fine spectral structure around 793‑810 nm becomes clearly resolved, indicating reduced phonon broadening and enhanced population stability of the Stark sublevels within the $^3H_4 \rightarrow {}^3H_6$ transition manifold [27]. In contrast, at room temperature, phonon-assisted relaxation processes lead to broader and partially merged emission peaks. This temperature-dependent evolution highlights the role of lattice vibrations in modulating the crystal-field splitting and line broadening of $Tm^{3+}$ emission in the $CeO_2$ host.

The energy levels within the $^3H_4$ manifold of $Tm^{3+}$ are weakly split due to the crystal-field effect originating from the cubic fluorite structure of $CeO_2$, in which $Tm^{3+}$ ions occupy highly symmetric lattice sites. Given the total angular-momentum quantum number J = 4, the $^3H_4$ energy level can split into a maximum of 2J + 1 = 9 Stark sublevels. However, the PL spectra reveal the presence of five distinct emission peaks, each corresponding to a different energy level.

Figure 4(c) presents the two-dimensional photoluminescence (PL) excitation map of the $Tm^{3+}$:$CeO_2$/Si film recorded at 4 K, where the excitation wavelength was systematically tuned while monitoring emission in the infrared region. Two intense and persistent emission bands centered at approximately ~1107 nm and ~1168 nm are observed across all excitation wavelengths, indicating transitions that are efficiently populated under various excitation conditions. In contrast, a distinct set of six weaker emission peaks at ~ 1205 nm, ~ 1210 nm, ~ 1406 nm, ~ 1464 nm, ~ 1492 nm, and ~ 1542 nm emerges only when the excitation wavelength lies within the narrow range of ~ 793 – 796 nm. This resonant behavior demonstrates that these infrared transitions are selectively driven through excitation of the $^3H_4$ manifold of $Tm^{3+}$ ions, corresponding to the $^3H_6 \rightarrow {}^3H_4$ absorption around ~ 793 nm. Based on their spectral positions, the peaks at ~ 1205 nm and ~ 1210 nm are attributed to the $^3H_5 \rightarrow {}^3H_6$ transition, while the ~ 1464 nm feature is assigned to the $^3H_4 \rightarrow {}^3F_4$ transition. The remaining bands at ~ 1406 nm, ~ 1492 nm, and ~ 1542 nm are interpreted as satellite Stark components associated with crystal-field splitting of these manifolds. Notably, excitation near ~ 808 nm, despite being in proximity to a strong PL band, does not produce additional infrared features beyond the two dominant peaks near ~ 1150 nm.

To further probe excitation-dependent emission behavior, up-conversion photoluminescence measurements were performed by tuning the excitation wavelength in the infrared range from 1135 nm to 1250 nm while monitoring emission in the visible and near-infrared regions, as shown in Figure 4(d). Emission features near 800 nm appear when the excitation wavelength is tuned around ~ 1210 nm, along with another strong up-conversion intensity for excitation near 1150 nm. While the precise excitation mechanism remains under investigation, the resonant nature of the up-conversion signal at ~ 1210 nm points to $Tm^{3+}$ ion's role in the up-conversion process and the existence of long-lived excited states of the Tm ions.

Time-resolved photoluminescence lifetime measurements were performed using a time-correlated single-photon counting approach. The sample emission was collected using the optical setup and directed to a superconducting nanowire single-photon detector from Single Quantum. The detector output was connected to a Swabian Instruments Time Tagger, which recorded photon arrival times with high temporal resolution. A pulsed excitation laser was used to excite the sample. The repetition rate of the excitation pulses was controlled using an acoustic-optic modulator placed in the beam path. The AOM allowed precise adjustment of the effective pulse repetition rate by gating the laser pulses, which ensured that the interval between excitation pulses was longer than the measured decay time. This prevented pulse overlap and enabled accurate extraction of the photoluminescence lifetime from the recorded photon arrival histogram.

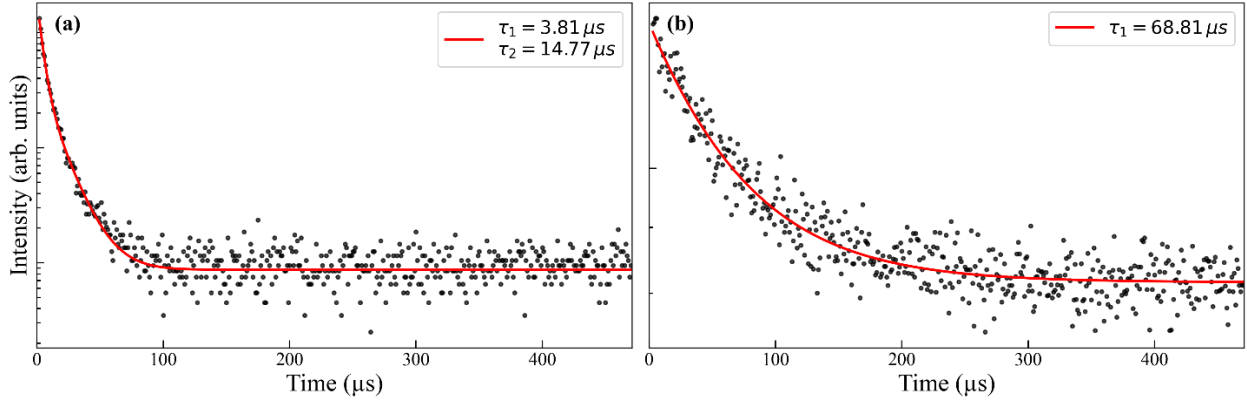

**Figure 5:** Lifetimes measurements on $Tm^{3+}$ transitions, **(a)** for $^3H_4 \rightarrow {}^3H_6$ excited using the up-conversion process using a 1210 nm laser. The bi-exponential decay curve is likely due to the mixing of two crystal-field sublevels at ~ 793 nm and ~ 808 nm. **(b)** Decay curve for $^3H_5 \rightarrow {}^3H_6$ transition excited using a 795 nm laser resonant with $^3H_4 \rightarrow {}^3H_6$ transition.

Under excitation at 1210 nm, an up-conversion process populates the $^3H_4$ level, leading to emission in the ~ 800 nm region. This emission is characterized by two distinct peaks at ~ 793 nm and ~ 808 nm. The emission was isolated using a 700 nm long-pass filter and an 850 nm short-pass filter. The resulting decay trace (Figure 5a) follows a bi-exponential decay profile, yielding a fast component of $\tau_1$ ~ 3.81 μs and a slower component of $\tau_2$ ~ 14.77 μs. This two-component decay is likely due to the different lifetimes of the two transitions at ~ 795 nm and ~ 808 nm. We found these lifetimes to be independent of the dopant concentration from 0.01-1%, indicating possible non-radiative processes.

The lifetime of the 1210 nm transition ($^3H_5 \rightarrow {}^3H_6$) was measured directly by exciting it through the $^3H_4 \rightarrow {}^3H_6$ transition. This emission was isolated using a specialized spectral filtering setup, providing a narrow ~ 3 nm bandwidth to ensure the purity of the signal from the 1210 nm level. As shown in Fig. 5b, the 1210 nm decay is well-described by a single-exponential model with a lifetime of $\tau_1$ ~ 68.81 μs. This significantly longer lifetime roughly 5 to 20 times longer than the 800 nm decay components confirms that the 1210 nm level ($^3H_5$) serves as a stable intermediate reservoir. Its long-lived nature is essential for the up-conversion process, as it allows sufficient time for the second excitation step required to reach the $^3H_4$ state.

To more clearly elucidate the origin of the short Tm lifetimes in the $CeO_2$ host, we compare these lifetimes with those of Er-doped $CeO_2$ films grown under identical conditions. The photoluminescence properties of the $CeO_2$(200nm)/$Ce_{0.99}Er_{0.01}O_2$(150nm)/$CeO_2$(200nm) film grown on YSZ (001) was investigated under 1418 nm excitation. The emission spectrum, shown in Figure 6a, exhibits two distinct peaks centered at ~ 1530 nm and ~ 1535 nm, corresponding to transitions from the Stark-split $^4I_{13/2}$ excited state to the $^4I_{15/2}$ ground state manifold. To isolate these features, the emission was passed through the same high-resolution spectral filtering setup used for the 1210 nm $Tm^{3+}$ measurements, utilizing a narrow 3 nm bandwidth to ensure the purity of the individual transition signals.

Time-resolved measurements revealed distinct decay dynamics for these transitions, presented in Figure 6 (b & c). The 1534 nm emission exhibits a single-exponential decay with a lifetime of $\tau_1 \sim 5.32$ ms, and the 1530 nm emission exhibits a faster decay rate, characterized by a lifetime of $\tau_1 \sim 2.94$ ms. The observed variation in lifetimes between the 1530 nm and 1534 nm peaks indicates that these transitions originate from different Stark levels within the $^4I_{13/2}$ manifold, which are likely influenced by local site symmetry.

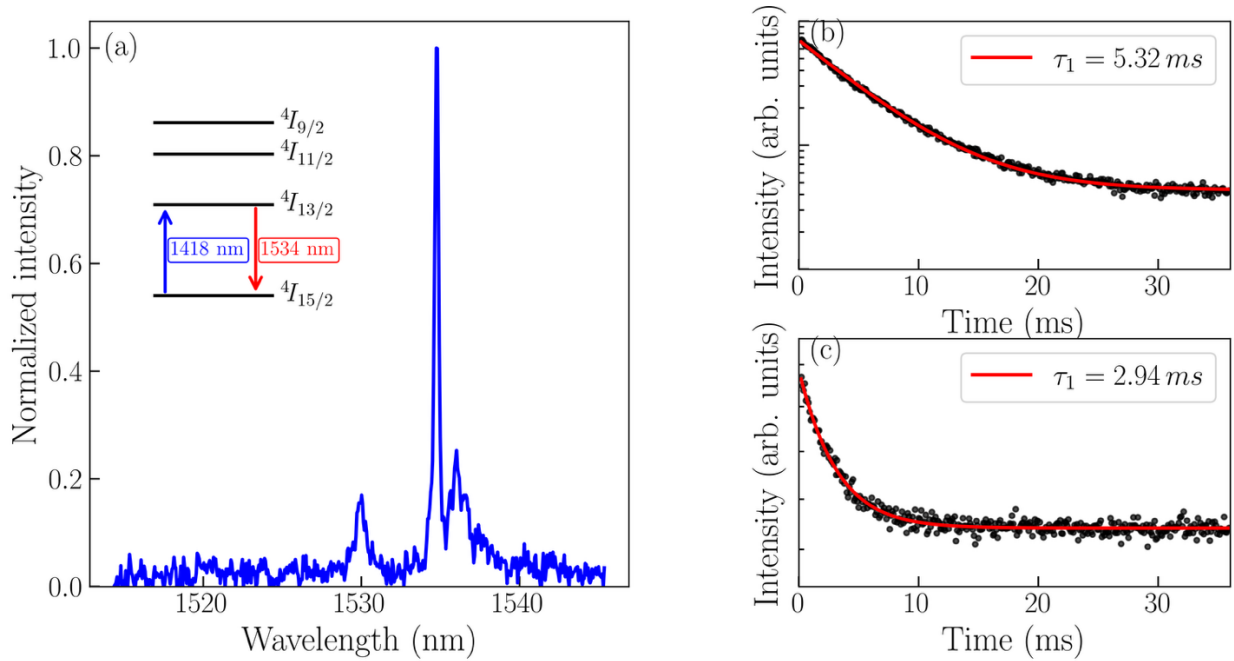

**Figure 6**: Photoluminescence and decay dynamics of 1 at.% Er:CeO$_2$ films grown on YSZ. **(a)** Emission spectrum under 1418 nm excitation, showing primary peaks at ~ 1530 nm and ~ 1535 nm. **(b)** Single-exponential decay fit for the 1534 nm transition, yielding $\tau_1 \sim 5.32$ ms. **(c)** Single-exponential decay fit for the 1530 nm transition, yielding $\tau_1 \sim 2.94$ ms. Spectral isolation for these measurements was achieved using the same filtering setup employed for the 1210 nm Tm$^{3+}$ measurements, incorporating a narrow 3 nm spectral selection bandwidth.

The relatively longer lifetimes observed for 1 at.% Er in the CeO$_2$ host are noteworthy, as comparable lifetimes have previously been reported for MBE grown Er-doped CeO$_2$ films with concentrations < 0.01%. This behavior may be associated with a reduced density of oxygen vacancies in the present films, which can influence the nonradiative recombination pathways and thereby impact the excited-state lifetime.

To identify whether a host-specific electronic-structure mechanism contributes to the difference in the lifetimes of Tm and Er, spin-polarized DFT+U calculations were performed by substituting a single Ce with Tm or Er in a 2×2×2 fluorite supercell (96 atoms, ~ 3% doping), and the geometry was fully relaxed. The atom-projected density of states (PDOS) is shown in Figure 7, with contributions from the rare-earth 4$f$ (blue), O 2$p$ (red), and Ce 4$f$ (green) orbitals resolved separately. The valence band maximum (VBM) of pristine CeO$_2$ is set to 0 eV.

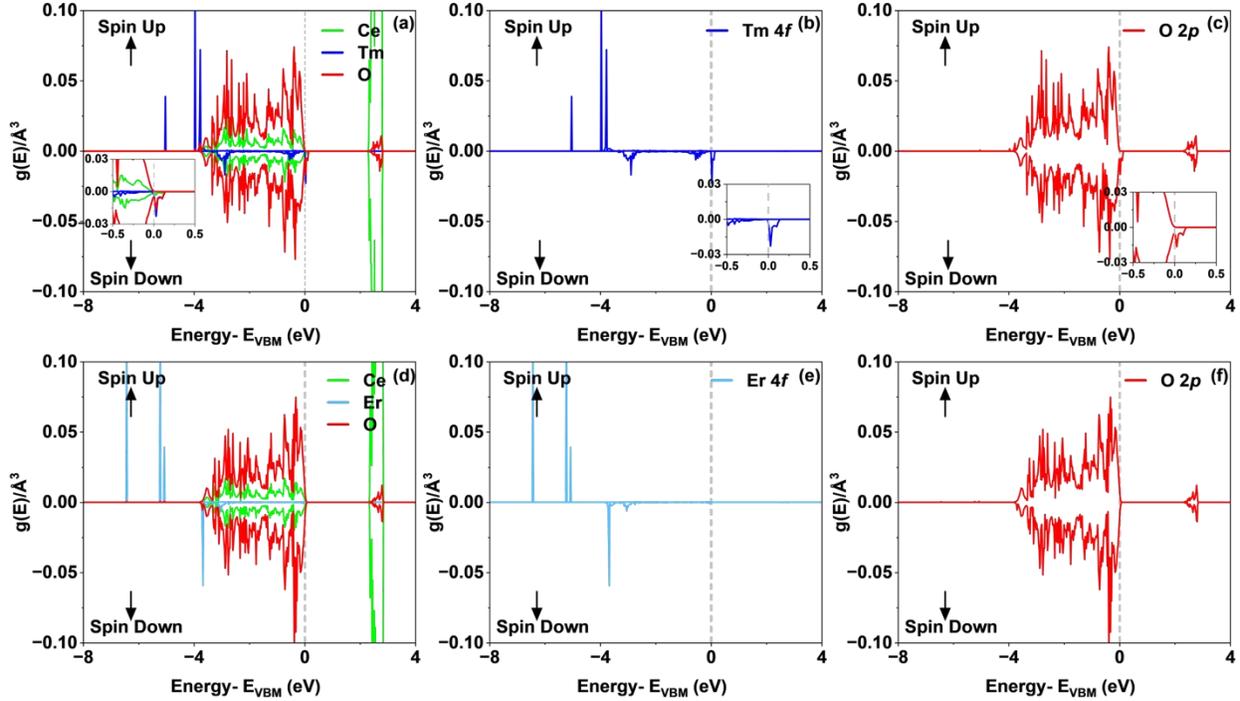

**Figure 7:** Spin-polarized density of states for (a) various elements in Tm-Doped $CeO_2$, (b) Tm $4f$ orbital in Tm-Doped $CeO_2$, (c) O $2p$ orbital in Tm-Doped $CeO_2$, (d) various elements in Er-Doped $CeO_2$, (e) Er $4f$ orbital in Er-Doped $CeO_2$, and (f) O $2p$ orbital in Er-Doped $CeO_2$. The top of the valence band maximum (VBM) from pristine of $CeO_2$ is set to 0 eV.

For the Er-doped system (Figure 7, bottom), the Er $4f$ manifold lies deep within the valence band, with the occupied states centered between approximately −8 and −4 eV below the VBM. At these energies the O $2p$ band, which carries its principal spectral weight between −4 and 0 eV, has negligible density. Because the Er $4f$ and O $2p$ states are not co-energetic, they do not mix: the $4f$ peaks remain narrow and atomic-like, preserving the standard picture of $4f$ electrons shielded by the closed $5s^25p^6$ subshell [43, 44]. No unoccupied Er-derived states appear near the VBM or within the host band gap.

The Tm-doped system (Figure 7, top) presents a different electronic structure. The occupied Tm $4f$ manifold sits at higher energies within the valence band relative to Er, with spin-up states between −6 and −4 eV. Several spin-down $4f$ components extend above −4 eV, entering the energy window where the O $2p$ band carries substantial spectral weight. Where the Tm $4f$ and O $2p$ states become co-energetic, the sharp atomic-like peaks characteristic of isolated $4f$ levels give way to broadened features, revealing the covalent hybridization between the dopant and host states. In addition, an unoccupied spin-down state appears just above the VBM, and real-space charge density visualization shows its wavefunction extends mainly onto neighboring O sites with mixed Tm $4f$ + O $2p$ character (Figure SI).

The site-projected magnetic moments also provide a measure of the hybridization contrast. For $Er^{3+}$ ($4f^{11}$), the Er site carries 2.97 $\mu_B$ with negligible spin density on the neighboring oxygen sublattice, confirming that the $4f$ spin density remains fully localized on the rare-earth site. For $Tm^{3+}$ ($4f^{12}$), the Tm site carries 2 $\mu_B$ (the expected Hund value for 12 electrons in the $4f$ shell) while

~ 1 $\mu_B$ of spin density resides on the neighboring oxygen atoms. This delocalization of one full Bohr magneton onto the sublattice directly quantifies the covalent Tm $4f$ – O $2p$ mixing and confirms that the $5s^25p^6$ shielding that normally isolates $4f$ electrons from the host is weakened for Tm in $CeO_2$.

Collectively, these rare-earth-doped systems highlight the critical role of the host lattice in mediating radiative and non-radiative relaxation processes. By contrasting the long-lived nature of the Er metastable states with the complex, fast-decaying up conversion channels of Tm, we establish a comprehensive experimental foundation for understanding dopant-host interactions. This data sets the stage for future theoretical modeling to quantify the crystal field parameters and electronic density of states that govern these distinct lifetime signatures.

## IV. Conclusions

In summary, we have investigated $CeO_2$ as a host system for quantum emitters with a focus on rare-earth ions. We have demonstrated epitaxial growth of crystalline Tm and Er doped $CeO_2$ films on YSZ (001), and Si (001) substrates. We have observed efficient photoluminescence from Tm ions, along with an up-conversion signal. Short excited-state lifetimes are observed in Tm-doped $CeO_2$ films, whereas comparatively longer lifetimes are measured in Er-doped $CeO_2$ films. This difference arises from the distinct interaction between the rare-earth dopants and the $CeO_2$ host lattice, particularly through the hybridization of RE $4f$ orbitals with O $2p$ states, as corroborated by first-principles DFT calculations. These findings provide important insight into the role of dopant-host electronic interactions and suggest a viable strategy for selecting appropriate rare-earth dopants as quantum sources in the non-magnetic $CeO_2$ host matrix for potential quantum applications. Another important result to be investigated further may be the role of oxygen vacancies in reducing the PL lifetimes which may explain the difference in the dopant density dependence of the lifetimes in MBE and PLD grown films.

## Acknowledgements

This material is based upon work supported by the U.S. Department of Energy, Office of Science, Office of Basic Energy Sciences under Award Number E-SC0025486. This award supported the thin film growth, X-ray-based characterization, and photoluminescence measurements. We gratefully acknowledge fruitful discussions with Mahdi Hosseini.

**References**

1. Jelezko, F. and J. Wrachtrup, *Single defect centres in diamond: A review.* physica status solidi (a), 2006. **203**(13): p. 3207-3225.
2. Childress, L. and R. Hanson, *Diamond NV centers for quantum computing and quantum networks.* MRS Bulletin, 2013. **38**(2): p. 134-138.
3. Hepp, C., et al., *Electronic Structure of the Silicon Vacancy Color Center in Diamond.* Physical Review Letters, 2014. **112**(3): p. 036405.

4. Sukachev, D.D., et al., *Silicon-Vacancy Spin Qubit in Diamond: A Quantum Memory Exceeding 10 ms with Single-Shot State Readout*. Physical Review Letters, 2017. **119**(22): p. 223602.
5. Evans, R.E., et al., *Photon-mediated interactions between quantum emitters in a diamond nanocavity*. Science, 2018. **362**(6415): p. 662-665.
6. Koehl, W.F., et al., *Room temperature coherent control of defect spin qubits in silicon carbide*. Nature, 2011. **479**(7371): p. 84-87.
7. Christle, D.J., et al., *Isolated Spin Qubits in SiC with a High-Fidelity Infrared Spin-to-Photon Interface*. Physical Review X, 2017. **7**(2): p. 021046.
8. Esmann, M., S.C. Wein, and C. Antón-Solanas, *Solid-State Single-Photon Sources: Recent Advances for Novel Quantum Materials*. Advanced Functional Materials, 2024. **34**(30): p. 2315936.
9. Graf, F.R., et al., *Photon-echo attenuation by dynamical processes in rare-earth-ion-doped crystals*. Physical Review B, 1998. **58**(9): p. 5462-5478.
10. Seo, H., et al., *Designing defect-based qubit candidates in wide-gap binary semiconductors for solid-state quantum technologies*. Physical Review Materials, 2017. **1**(7): p. 075002.
11. Zhang, G., et al., *Material platforms for defect qubits and single-photon emitters*. Applied Physics Reviews, 2020. **7**(3).
12. Somjit, V., et al., *An NV− center in magnesium oxide as a spin qubit for hybrid quantum technologies*. npj Computational Materials, 2025. **11**(1): p. 74.
13. Johnson, J.M., et al., *Unusual Formation of Point-Defect Complexes in the Ultrawide-Band-Gap Semiconductor $\beta\text{-}\mathrm{Ga}_2\mathrm{O}_3$*. Physical Review X, 2019. **9**(4): p. 041027.
14. Bertaina, S., et al., *Rare-earth solid-state qubits*. Nature Nanotechnology, 2007. **2**(1): p. 39-42.
15. O'Brien, C., et al., *Interfacing Superconducting Qubits and Telecom Photons via a Rare-Earth-Doped Crystal*. Physical Review Letters, 2014. **113**(6): p. 063603.
16. Thiel, C.W., T. Böttger, and R.L. Cone, *Rare-earth-doped materials for applications in quantum information storage and signal processing*. Journal of Luminescence, 2011. **131**(3): p. 353-361.
17. Ortu, A., et al., *Storage of photonic time-bin qubits for up to 20 ms in a rare-earth doped crystal*. npj Quantum Information, 2022. **8**(1): p. 29.
18. Böttger, T., et al., *Spectroscopy and dynamics of $\mathrm{Er}^{3+}:\mathrm{Y}_2\mathrm{SiO}_5$ at $1.5\,\mu\mathrm{m}$*. Physical Review B, 2006. **74**(7): p. 075107.
19. Williams, D.K., et al., *Preparation and Fluorescence Spectroscopy of Bulk Monoclinic Eu3+:Y2O3 and Comparison to Eu3+:Y2O3 Nanocrystals*. The Journal of Physical Chemistry B, 1998. **102**(6): p. 916-920.
20. Serrano, D., et al., *Coherent optical and spin spectroscopy of nanoscale $\mathrm{Pr}^{3+}:\mathrm{Y}_2\mathrm{O}_3$*. Physical Review B, 2019. **100**(14): p. 144304.

21. Stouwdam, J.W. and F.C.J.M. van Veggel, *Near-infrared Emission of Redispersible Er3+, Nd3+, and Ho3+ Doped LaF3 Nanoparticles.* Nano Letters, 2002. **2**(7): p. 733-737.
22. de Leon, N.P., et al., *Materials challenges and opportunities for quantum computing hardware.* Science, 2021. **372**(6539): p. eabb2823.
23. Burkard, G., et al., *Semiconductor spin qubits.* Reviews of Modern Physics, 2023. **95**(2): p. 025003.
24. Kanai, S., et al., *Generalized scaling of spin qubit coherence in over 12,000 host materials.* Proceedings of the National Academy of Sciences, 2022. **119**(15): p. e2121808119.
25. Inaba, T., et al., *Epitaxial growth and optical properties of Er-doped CeO2 on Si(111).* Optical Materials Express, 2018. **8**(9): p. 2843-2849.
26. Grant, G.D., et al., *Optical and microstructural characterization of Er3+ doped epitaxial cerium oxide on silicon.* APL Materials, 2024. **12**(2).
27. Kramida, A., Ralchenko, Yu., Reader, J., and NIST ASD Team (2024), *NIST Atomic Spectra Database*, G. National Institute of Standards and Technology, Editor. 2024.
28. Parashar, M., et al., *Probing elemental diffusion and radiation tolerance of perovskite solar cells via non-destructive Rutherford backscattering spectrometry.* APL Energy, 2024. **2**(1).
29. Kresse, G. and J. Hafner, *Ab initio molecular dynamics for liquid metals.* Physical Review B, 1993. **47**(1): p. 558-561.
30. Kresse, G. and D. Joubert, *From ultrasoft pseudopotentials to the projector augmented-wave method.* Physical Review B, 1999. **59**(3): p. 1758-1775.
31. Blöchl, P.E., *Projector augmented-wave method.* Physical Review B, 1994. **50**(24): p. 17953-17979.
32. Perdew, J.P., K. Burke, and M. Ernzerhof, *Generalized Gradient Approximation Made Simple.* Physical Review Letters, 1996. **77**(18): p. 3865-3868.
33. Dudarev, S.L., et al., *Electron-energy-loss spectra and the structural stability of nickel oxide: An LSDA+U study.* Physical Review B, 1998. **57**(3): p. 1505-1509.
34. Castleton, C.W.M., J. Kullgren, and K. Hermansson, *Tuning LDA+U for electron localization and structure at oxygen vacancies in ceria.* The Journal of Chemical Physics, 2007. **127**(24).
35. Larson, P., et al., *Electronic structure of rare-earth nitrides using the $\mathrm{LSDA}+U$ approach: Importance of allowing $4f$ orbitals to break the cubic crystal symmetry.* Physical Review B, 2007. **75**(4): p. 045114.
36. Sanna, S., et al., *Rare-earth defect pairs in GaN: $\text{LDA}+U$ calculations.* Physical Review B, 2009. **80**(10): p. 104120.
37. Denawi, H., P. Karamanis, and M. Rérat, *Electronic and magnetic properties of yttria-stabilized zirconia (6.7 mol% in Y2O3) doped with Er3+ ions from first-principle computations.* Journal of Materials Science, 2021. **56**(13): p. 8014-8023.
38. Monkhorst, H.J. and J.D. Pack, *Special points for Brillouin-zone integrations.* Physical Review B, 1976. **13**(12): p. 5188-5192.
39. Shi, Y., et al., *Growth of Highly Strained CeO2 Ultrathin Films.* ACS Nano, 2016. **10**(11): p. 9938-9947.

40. Skorodumova, N.V., M. Baudin, and K. Hermansson, *Surface properties of ${\mathrm{CeO}}_{2}$ from first principles.* Physical Review B, 2004. **69**(7): p. 075401.
41. Phoka, S., et al., *Synthesis, structural and optical properties of CeO2 nanoparticles synthesized by a simple polyvinyl pyrrolidone (PVP) solution route.* Materials Chemistry and Physics, 2009. **115**(1): p. 423-428.
42. Chrysochoos, J. and A.H. Qusti, *Cross-relaxation of the 1G4 state of Tm3+ in POCl3:SnCl4.* Journal of the Less Common Metals, 1989. **148**(1): p. 253-257.
43. Dieke, G.H. and H.M. Crosswhite, *The Spectra of the Doubly and Triply Ionized Rare Earths.* Applied Optics, 1963. **2**(7): p. 675-686.
44. Carnall, W.T., et al., *A systematic analysis of the spectra of the lanthanides doped into single crystal LaF3* The Journal of Chemical Physics, 1989. **90**(7): p. 3443-3457.